\documentstyle[NATO,numreferences]{CRCKAPB} 

\newcommand{\cago}{$^{12}{\rm C}(\alpha,\gamma)^{16}{\rm O}$\ }

\begin{opening}

\title{WHITE DWARF SEISMOLOGY AND THE \cago RATE}

\author{TRAVIS S. METCALFE}
\institute{Theoretical Astrophysics Center, Aarhus University\\
           Ny Munkegade bldg. 520, 8000 Aarhus C, DENMARK} 

\end{opening}

\begin{document}

\begin{abstract}
Recent determinations of the internal composition and structure of 
two helium-atmosphere variable white dwarf stars, GD~358 and CBS~114,
have led to conflicting implied rates for the \cago reaction. If we
assume that both stars were formed through single-star evolution,
then the initial analyses of their pulsation frequencies must have
differed in some systematic way. I present improved fits to the two
sets of pulsation data, helping to resolve the tension between the 
initial results.
\end{abstract}

\vskip -20pt
\section{Introduction}

When a white dwarf is being formed in the core of a red giant star during
helium burning, there are two nuclear reactions that compete for the
available helium nuclei: the 3$\alpha$ reaction, which combines three
helium nuclei to form carbon, and the \cago reaction, which combines an
additional helium nucleus with the carbon to form oxygen. At a given core
temperature and density, the relative rates of these two reactions largely
determines the C/O ratio in the resulting white dwarf star. The rate of
the 3$\alpha$ reaction is known to about 10\% precision, but the \cago
reaction is still uncertain by about 40\%. So, if we can measure the C/O
ratio in the core of a pulsating white dwarf, it is effectively a
measurement of the \cago reaction rate. The C/O ratio is interesting by
itself, since the core composition in our models affect the derived
cooling ages of white dwarfs by up to a few Gyr \cite{1}. But we can
also use it to provide an independent measurement of a nuclear reaction
that is important to many areas of astrophysics, from the energetics of
type Ia supernovae explosions to galactic chemical evolution.

The model-fitting method that I describe below has only been applied to DB
white dwarfs, since they are structurally the simplest. But in principle
it can be extended to the pulsating DA stars quite easily, and with a
little more work to the DOVs. Presently, the method requires that the
spherical degree of the pulsation modes is known, and sufficient data
exist for only two stars---though we have just finished a Whole Earth
Telescope run on a third object. The first application was to GD~358,
which showed 11 consecutive radial overtones with the same spherical
degree during a WET run in 1990 \cite{2,3}. The second application was
only recently finished, and came from single-site data on the star
CBS~114, which showed 7 independent modes that all appear to be $\ell$=1
\cite{4}. What we set out to do was search for a theoretical model that
could reproduce, as closely as possible, the pulsation periods that we
have observed in these stars.

\section{Model-Fitting}

We adjusted five different model parameters to try to match the observed
periods. To make the final result as objective as possible, we wanted to
explore the broadest range for each model parameter, defining the limits
of the search based only on the physics of the model and on observational
constraints. We allowed the mass to be anywhere from 0.45 to 0.95
M$_{\odot}$, which encompasses the vast majority of known white dwarf
masses \cite{5}. The temperature of our models was allowed to vary from
20,000 to 30,000~K, which easily includes the spectroscopic temperature
determinations of all of the known DBV white dwarfs whether or not trace
amounts of hydrogen are included in their atmospheres \cite{6}. We looked
at helium layer masses ranging from a fractional mass of 10$^{-2}$ where
helium burning will begin at the base of the envelope, down to a few times
10$^{-8}$, close to the limit where our models no longer pulsate \cite{7}.
The final two parameters describe a simple C/O profile that has a constant
ratio ($X_{\rm O}$) out to some fractional mass point ($q$), where it then
decreases linearly in mass to zero oxygen at the 95\% mass point. The
important features from the standpoint of pulsations are the central C/O
ratio and the location and slope of the composition transition.

To get reasonable resolution, we allowed 100 values for each parameter
inside these limits, so there are 10$^{10}$ possible combinations of these
five model parameters. Even if we had 1000 of today's fastest processors,
it would still take more than a year to calculate all of these models, so
we used a slightly more clever method employing a genetic algorithm to
explore the many possibilities. The way genetic algorithms work is
initially like a Monte Carlo method, where we just take a random sample of
parameter combinations. After this initial random sampling, a genetic
algorithm explores new regions of the search space based on a sort of
survival of the fittest scheme. By passing simulated data through this
process, we can quantify how long we need to let it run to find the
correct set of parameters most of the time, and by running the entire
process several times with different random initialization, we can ensure
a very high probability of finding the globally optimal set of model
parameters. In the end, the method requires a few million models to be
calculated, concentrated mostly around the regions of the search space
that yield better than average matches to the observations. So in addition
to the optimal set of model parameters, we also end up with a fairly
decent map of the search space, which gives us some sense of the
uniqueness of the final answer.

To learn what the optimal values of the mass and central C/O ratio say
about the \cago reaction, we need additional models that evolve a star
from the main sequence through the red giant phase and into a white dwarf
\cite{8}. For a given white dwarf mass, there are several things in
the models that can be adjusted to change the central C/O ratio, but the
ingredient that affects it the most is the \cago reaction. So, all we have
to do is adjust this rate in the models until we end up with the mass and
central C/O ratio that matches the fit from the genetic algorithm.

\section{Results}

In the initial application of this method to the data from the two stars,
we found that GD~358 implied a reaction rate significantly higher than the
extrapolations from laboratory measurements \cite{9}, while CBS~114 was
right in line with the expectations \cite{4}. This led us to speculate
that there might be some systematic error affecting our analysis of the
two stars in different ways. There was another slight worry in the initial
results: the masses and temperatures for both stars differed significantly
from those inferred from spectroscopy. We thought this may have resulted
from the use of slightly different mixing length parameters than the
spectroscopic studies, so we repeated the fits using ML2/$\alpha$=1.25 to
see if the discrepancy would disappear. In doing so, we also realized that
there {\it was} a systematic difference in the way we were analyzing the
two stars: for GD~358 the pulsation modes were consecutive radial overtones, 
so we were using both the periods and the spacings between the periods to 
judge which models provided the best match. For CBS~114 there was a gap in 
the sequence of observed modes, so we only used the periods themselves. When
we repeated the fit for GD~358, we used only the periods to determine the
best fit, just as we had for CBS~114.

\begin{table}[htb]
\begin{center}
\caption{Optimal ML2/$\alpha$=1.25 Models}
\begin{tabular}{lccccccc}
\hline
Object & $T_{\rm eff}$ & $M/M_{\odot}$ & $\log(M_{\rm He}/M_*)$ &
$X_{\rm O}$ & $q$ & $\sigma_{\rm P}(s)$ & $S_{300}$ \\
\hline
GD~358  & 21,300 & 0.695 & $-$2.95 & 0.69 & 0.49 & 1.11 & $215\pm20$ \\
CBS~114 & 20,500 & 0.745 & $-$6.77 & 0.58 & 0.51 & 0.43 & $160\pm20$ \\
\hline
\end{tabular}
\end{center}
\end{table}

The results of the new fits are shown in Table 1. Notice that both objects
yield a rate for the \cago reaction that is consistent with laboratory
extrapolations ($S_{300}=200\pm80$, \cite{10}), but GD~358 still seems to
be a bit high compared to CBS~114, so this is only part of the answer.
Also note that switching to the mixing length parameters used in the
spectroscopic analysis did not resolve the differences between the implied
masses and temperatures: the two methods of inferring these parameters do
not agree. Curiously, the optimal model of GD~358 has a thick helium
layer, but for CBS~114 it is thin. Finally, both models show the
transition point from a constant C/O ratio near the same fractional mass,
and this location does not favor convective overshoot. Probably the
decisive tests of all of these puzzles will come as we apply this method
to additional DBV white dwarfs.

\section{Conclusions}

By measuring the interior composition of pulsating white dwarfs, we can
get precise measurements of the important \cago reaction rate, and as our
models improve we can be more and more sure that they are not only
precise, but also accurate. The model-fitting tool that we have used,
involving a genetic algorithm, is a very powerful way to explore large
ranges of interesting physical parameters and find the globally optimal
model to match the observations. And since it evaluates so many models
along the way, it produces some good maps of the search space in the
regions where it is most interesting. True, this method is still
computationally intensive, but there is no getting around that if we want
the global solution, and Linux clusters are getting cheaper and cheaper.
Finally, in the future we hope to be able to say something more about the
detailed shape of the C/O profile all the way from the center to the
surface.


\begin{thebibliography}{}

\bibitem{1} Fontaine, G., Brassard, P., \& Bergeron, P. 2001,
{\it PASP}, {\bf 113}, 409.

\bibitem{2} Winget, D. E.~et al. 1994, {\it ApJ}, {\bf 430}, 839.

\bibitem{3} Metcalfe, T. S., Winget, D. E., \& Charbonneau, P. 2001,
{\it ApJ}, {\bf 557}, 1021.

\bibitem{4} Handler, G., Metcalfe, T. S., \& Wood, M. A. 2002, 
{\it MNRAS}, in press.

\bibitem{5} Napiwotzki, R., Green, P. J. \& Saffer, R. A. 1999,
{\it ApJ}, {\bf 517}, 399.

\bibitem{6} Beauchamp, A.~et al. 1999, {\it ApJ}, {\bf 516}, 887.

\bibitem{7} Bradley, P. A. \& Winget, D. E. 1994a, {\it ApJ}, 
{\bf 421}, 236.

\bibitem{8} Salaris, M., et al. 1997, {\it ApJ}, {\bf 486}, 413.

\bibitem{9} Metcalfe, T. S., Salaris, M. \& Winget, D. E. 2002,
{\it ApJ}, {\bf 573}, 803.

\bibitem{10} Angulo, C.~et al. 1999, {\it Nuc.~Phys.~A}, {\bf 656}, 3.

\end{thebibliography}
\end{document}